# Switching Behavior of Bulk, Fast Ion Conducting, Vitreous AgI-Ag$_2$O-MoO$_3$ Solids with Inert Electrode

B. Tanujit[a], G. SreevidyaVarma[b] and S. Asokan[a,*]

[a] *Department of Instrumentation and Applied Physics, Indian Institute of Science, Bangalore – 560 012, India*
[b] *Department of Physics, Presidency University, Bangalore – 560 064, India*

## Abstract

Developing efficient, fast performing and thermally stable Silver iodide based fast ion conducting solids are of great interest for resistive switching applications, but still remain challenges. Metallization in bulk, behavior of threshold voltage profile over composition and corrosion reactions are few of these challenges. In this work, the switching behavior of bulk, fast ion conducting, vitreous (AgI)$_x$-(Ag$_2$O)$_{25}$-(MoO$_3$)$_{75-x}$, for $60 \leq x \leq 40$ solids, has been investigated, in order to understand the switching mechanism with theinert electrodes. By using inert electrodes, the switching becomes irreversible, memory type. The switching mechanism is electrochemical metallization process. The inert electrodes restrain ionic mass transfer but exhibit low barrier to electron transfer allowing the cathodic metallization reaction to reach Nernst equilibrium faster. Cations involved in this process transport thorough the free volume within the solid structure and follows Mott-Gurney model for electric field driven thermally activated ion hopping conductivity model. This model along with the thermal stability profile provide a narrow region within composition with better switching performance based on swiftness to reach threshold voltage and less power loss. Traces of anionic contribution to metallization are absent. Moreover, anodic oxidation involves reactions that cause bubble formation and corrosion.

---

[*] Corresponding Author email: sundarrajan.asokan@gmail.com, sasokan@iisc.ac.in

Fax: 91-80-23608686; Phone. +91-80-22933195, 91-80-22932271

## 1. Introduction

Fast ion conducting (FIC) vitreous solids have been investigated for nearly three decades for their high ionic conductivity ($10^{-5}$-$10^{-2}$ $\Omega^{-1}$ cm$^{-1}$) in solid phase at room temperature[1]. Various characterization techniques and theoretical studies have been conducted to understand the formation, ionic transport, structure and consequently several models for ion transport mechanism have been proposed for these solids. In addition, efforts have been made to exploit these materials for different applications such as solid state electrolytes in batteries, ambient temperature oxygen sensor, chemical sensor, electron beam recording materials [2] etc. Meanwhile, B. Vaidhyanathan et al. reported [3-4] on the memory behavior of bulk silver iodide based FIC glasses with different glass formers and of varying thicknesses. However, the thickness dependence of threshold voltage ($V_{th}$) has not been fully understood in their work. Also, their interpretation of switching mechanism differs considerably from the conventional perspective. For the metal-insulator-metal (MIM) structured ionic memories, electrochemically active metal anode material gets oxidized and dissolves into the solid electrolyte and reduces to cathode, forming metallic filament [5-6]. Vaidhyanathan et al. have suggested a probable mechanism that consists of both cathodic reduction of $Ag^+$ ion and anodic oxidation of $[MoO_4]^{2-}$ ion, along with both cationic and anionic transport.

The present work primarily focuses on the understanding of the memory behavior of the bulk AgI-$Ag_2O$-$MoO_3$ vitreous solids with different compositions, especially, on the role of the inert electrode in the memory behavior of the bulk electrolyte. The inert electrodes restrain ionic mass transfer but allow electrons to continue electrochemical reactions. This restriction to ionic mass transport may cause polarization effect near the electrodes and could instigate the filament formation. By performing cyclic voltammetry (CV) experiment, we look into the basis of metallization and distinguish it from a process of slow electron transfer reaction. The metallization involves electrochemical reduction of $Ag^+$ ions but the oxidation process involves reaction that results in corrosion in the interface. Reactions that are responsible for this interfacial oxidation have been understood. Furthermore, the vitreous state plays important role in conduction process because it involves defect based ion hopping and network structure formation that provides thermal stability. Superionic conductors with crystalline structure e.g. $RbAg_4I_5$ [7], $Ag_2S$ [8], $Cu_2S$ [9] exhibit different mechanism for these two processes. Role of this vitreous state has been understood and discussed from results obtained from differential scanning calorimetry (DSC) study. Mott-Gurney model for electric field driven, thermally activated ion hopping has been used to understand ion migration, thickness dependence and $V_{th}$ – composition (x) profile to classify the samples according to their switching behavior.

The electrical switching phenomenon is generally exemplified by a rapid transition from a high resistance OFF state (HRS) to a low resistance ON state (LRS). The HRS of the present samples is ~10 MΩ, and the LRS is ~1-2 Ω for a 0.2 mm thick sample. The results obtained indicate that these vitreous solids exhibit a nearly ideal, fast, memory type irreversible switching behavior with very low power dissipation. Unlike the MIM structured cation based ionic memories [6],a voltage sweep from positive to negative bias cannot alter the memory from LRS to HRS in these samples. The reason behind the irreversibility has been discussed in terms of the switching mechanism and electrode type.

## 2. Experimental

### A. Sample Preparation

Bulk$(AgI)_x$-$(Ag_2O)_{25}$-$(MoO_3)_{75-x}$, for $60 \leq x \leq 40$,vitreous solids have been prepared by microwave melting – metal plate quenching method. The fundamental idea behind AgI based solid electrolytes with high ionic conductivity at room temperature is to hinder the $\alpha \rightarrow \beta$ phase transition at lower temperatures, by introducing stabilizer ions into the lattice [10], i.e. preserving the $\alpha$-AgI at much lower temperature, in other words, a lower phase transition temperature [11]. This can be efficiently achieved by microwave irradiation technique, the $\beta$ — $\alpha$ phase transition occurs at around $100^o$C in this case [11, 12]. The microwave energy interacts with the crystal via phonon coupling and the low-lying transverse optic (TO) modes dominate; $Ag^+$ mostly moves in these low energy modes that leads to Frenkel defects in the structure, allowing ionic conduction [12].

Table 1 summarizes the mole percentage of the constituent materials. All the constituent materials have individual significance. $MoO_3$ is a conditional glass former, unlike other nonmetal oxides (e.g. $SiO_2$, $GeO_2$, $P_2O_5$ etc.), the pure liquid of $MoO_3$ essentially requires a network modifier (e.g. $Na_2O$, $Li_2O$, $Ag_2O$ etc.) to form glass [13]. The other importance of $MoO_3$is revealed later in this paper.The introduction of a dopant salt, e.g. AgI in the glassy matrix causes expansion of the matrix by forming free volume/pathways and thus enhances the ionic conductivity [14]. The compositions are planned in a way to understand the balance between matrix former and matrix expander, higherAgI concentration causes stable, non-conductive $\beta$-phase formation, AgI crystallization; while lower AgI concentration in the matrix causes less ionic conductivity, thermal instability and brittleness. Thus, for higher ionic conductivity and thermal stability, a balance in thecomposition is essential. Our present compositions surround the central composition $(AgI)_{50}(Ag_2O)_{25}(MoO_3)_{25}$ [1, 15-17], while keeping $Ag_2O$ molar concentration fixed  and changing the AgI/$MoO_3$ molar ratio proportionately.

| Sample Level | Mole percentage | | |
|---|---|---|---|
| | AgI | $Ag_2O$ | $MoO_3$ |
| S1 | 60 | 25 | 15 |
| S2 | 57.5 | 25 | 17.5 |
| S3 | 55 | 25 | 20 |
| S4 | 52.5 | 25 | 22.5 |
| S5 | 50 | 25 | 25 |
| S6 | 47.5 | 25 | 27.5 |
| S7 | 45 | 25 | 30 |
| S8 | 42.5 | 25 | 32.5 |
| S9 | 40 | 25 | 35 |

Table 1: Mole percentage of AgI, $Ag_2O$ and $MoO_3$

The required quantities of AgI (99.9%), $Ag_2O$ (99.9%) and $MoO_3$ (99.9%) are thoroughly mixed in an Agate stone mortar and placed in a silica crucible, inside a microwave oven, operating at 2450 MHz with a continuous power level of 900 watts maximum. After 10-12 minutes the yellow homogenous melt is quenched down to room temperature in between two stainless steel plates. As AgI is highly photosensitive theprepared samples have been well preserved in a steel container to avoid photo induced modifications.

### B. Material Characterizations

X-ray diffraction studies (XRD) have been conducted using a Bruker Powder diffractometer with $CuK_\alpha$ radiation in the $2\theta$ range from $10^o$ to $90^o$ at a rate of $10^o$/min. The XRD study confirms the amorphous nature of the samples (Supplementary Figure1). DSC studies have been performed using a Mettler Toledo 822e ADSC instrument with $STAR^e$ software. Thermal scans are done in the temperature range $40^o$C to $250^o$C at a rate of $3^o$C/min for samples ~15 mg of approximately equal thickness, Argon is used as the purge gas at a flow rate of 75ml/min. Raman spectroscopic studies of the solid samples have been performed in a Horiba JobinYvon (LabRAM HR) spectrometer. The present samples are photosensitive to 514.5 nm Argon laser. The exposure leaves a permanent visibly dark spot on the sample which is basically crystallization due to laser heating [18]. In order to avoid that, a near IR (NIR) source of 784.8 nm diode laser is used. Scanning Electron Microscope (SEM) and Energy-dispersive X-ray spectroscopy (EDS) study across the vertical cross section of the metallic filament in the switched portion of a sample has been performed in a high resolution ULTRA 55-GEMINI SEM. CV measurements has been carried out in a CHI660C electrochemical workstation. The maximum ($E_{max}$) and minimum ($E_{min}$) voltages are kept at 0.4 and -0.4 V respectively, with scan rate of 0.1 V/s and sensitivity 1e-5 A/V.

### C. Electrical Switching Experiment

The electrical switching experiments have been carried out using a Keithley source meter (model 2410) controlled by Lab VIEW 6i (National Instruments). The source meter is capable of sourcing current in the range of 0-21 mA at a maximum compliance voltage of 1100 V. Samples carefully polished to 0.20 mm thickness using 400 and 4000 grade emery sheet, are placed in between a spring loaded point contact top electrode and a flat plate bottom electrode made of brass(Supplementary Figure 2).

### 3. Results and Discussions
#### A. Thermal Characterizations; Glass Transition Temperature ($T_g$) and Composition

For any potential solid state device (SSD) material, scaling is an important factor to improve density, performance, power consumption and cost effectiveness. Apparently, higher scaling factor results in larger data size that reaches overheating point quicker, causing device malfunctioning, lack of data persistency and data loss. Thermal stability is also essential, since its absence sets aside any commercial usefulness of $RbAg_4I_5$ though a fast ion conductor [19]. Moreover, during switching, local temperature enhancement can cause uncontrolled damage to the system. Thus, to attain chip level reliability, thermal characterization is an important concern not only in chip level but also in the bulk material. DSC is a sensitive, non-perturbing thermo-analytic technique, primarily used to determine characteristic temperatures e.g. crystallization ($T_x$), melting ($T_m$), glass transition temperature ($T_g$) etc. and energetics of phase transition, thermal stability, conformational changes and to investigate thermodynamic properties [20].

Figure 1 shows the DSC thermograms of samples (from up to down) having a gradual decrement in AgI and increment in $MoO_3$ concentration. In general, all of these vitreous solid samples exhibit glass transition and upon heating beyond $T_g$, the supercooled liquid state relaxes to thermodynamically stable crystalline structure [21]. S1,S2 andS3, samples with high AgI concentration exhibit two exothermic and two endothermic peaks, with $T_g$ of 42.5°C, 48°C and 52°C respectively. The first exotherm at 83°C for S1 is due to AgI crystallization [22] that shifts towards the second exothermic peak while lowering the AgI concentration. The first endotherm at ~ 152°C is assigned to β → α transition of AgI [22]. Following this crystallization, cation jump frequency drops and hence conductivity decreases [22]. Besides, higher temperature causes Frenkel defect formation in AgI, subsequent avalanche and sub-lattice melting. From S4 to S9, the compositions with relatively lower AgI concentration do not display the exotherm for AgI crystallization and the endotherm for β → α transition.  In the early studies on devitrification, metastability and phase diagram of the pseudo-binary system $(AgI)_{1-x}(Ag_2MoO_4)_x$ $(0.1 \leq x \leq 0.9)$, the second exotherm at ~ 110°C has been assigned to the crystallization of the metastable compound 2AgI-$Ag_2MoO_4$ (2:1), and the second endotherm at ~ 180°C is due to the decomposition of the eutectic between AgI and the 2:1 compound [23].S3, S4 and S5 show a very small endothermic hump resulting an incongruent melting, caused by this eutectic impurity. The absorbance of this hump into a coherent melting temperature ($T_m$) signifies that the incongruence is steadily decreasing i.e. formation of a regular mixture of cations and anions.S7, S8 and S9 with higher $MoO_3$ concentration, show only one exotherm at ~ 157°C. These exotherms are sharper for S8 and S9. The sharpness is caused by fast phase transformation.

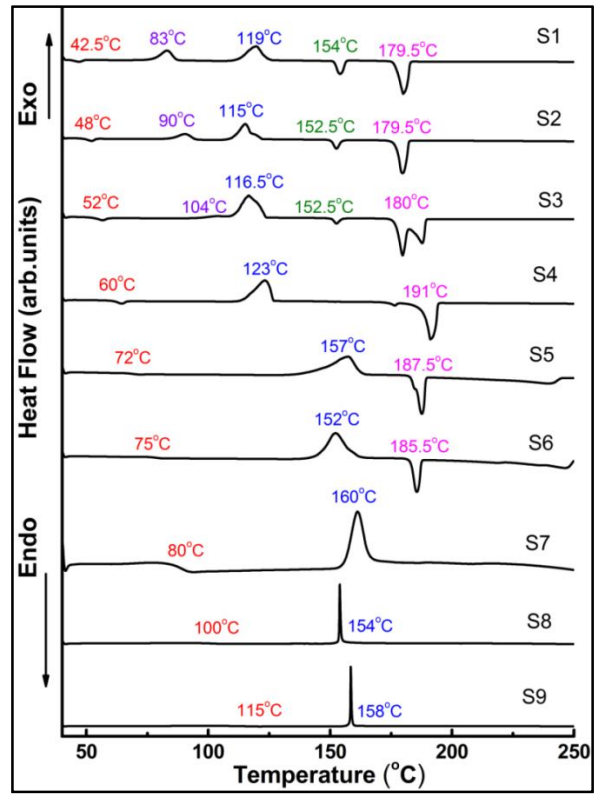

Figure 1: DSC thermogram of the samples; Heat flow (arbitrary units) vs. Temperature (°C)

Figure 2 shows the $T_g$ and stability ($T_x$ - $T_g$) profile with respect to composition. $T_g$ rises almost linearly with the $MoO_3$ concentration because of an increase in covalent Mo–O bond in the matrix which causes increased network connectivity. However, this does not ensure better thermal stability. The stability factor, $T_x$-$T_g$, indicates the devitrification tendency when heated above $T_g$. In the present results, the decreasing stability profile for S9 and S8 is interesting to note. As the network connectivity increases within the system, while the network essentially constitutes of covalent type bonds, a transition in conformational entropy takes place that results in stability downturn. This can be a matter of further research.

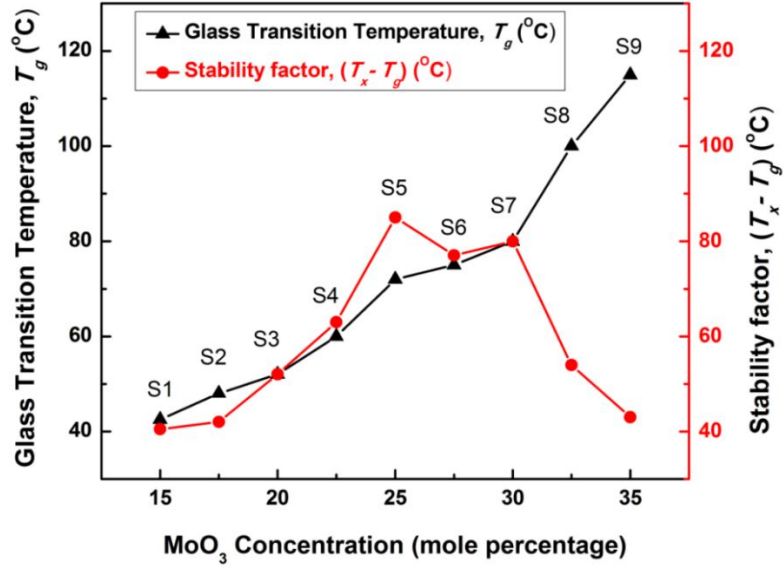

Figure 2: $T_g$ and stability factor ($T_x - T_g$) vs. MoO$_3$ concentration (mole percentage)

The plateau region from S7 to S5 corresponds to the maximum stability for these samples, indicating a higher glass forming ability. With increasing AgI concentration, stability factor decreases towards S1. This thermally stable region, for these vitreous solids corresponds to samples with congruent mixing of cations and anions, sufficiently ionic to exhibit fast switching and necessarily covalent to retain the atomic structure of supercooled liquidsduring thermal treatments. Hence the DSC study, on one hand classifies the composed samples based on their thermodynamic attributes and on the other, provides a preliminary insight into the structural counterpart of thermal behavior and composition.

**B. Memory type Electrical Switching Behavior**

The input direct current, which is a triangular pulse which increases from 0 to $I_{max}$ (for this particular study, $I_{max} = 1$ mA) and then decreases to 0 in steps of 0.01667 mA (approx.) with a time period of 0.125 seconds for each step, is passed through the sample (Figure 3a) and the voltage developed across the sample is measured simultaneously (Figure 3b). AgI-Ag$_2$O-MoO$_3$ normally has a very high resistance ($\sim$ 10 M$\Omega$) and can be considered to be in ON state. As the current increases gradually to the value of threshold current, $I_{th}$, the voltage across the electrodes increases to the maximum value (from A to B, Figure 3b) called threshold voltage, $V_{th}$. Beyond $I_{th}$, the voltage drops abruptly (from B to C, Figure 3b) and any further change in applied current cannot reshape the voltage profile significantly (from C to E, Figure 3b). In other words, the effective volume of the electrolyte that is in contact with the electrodes latches permanently at a very low resistance state. Unlike the RESET current impulse for Chalcogenide glasses [24] or a bipolar voltage sweep for MIM structured polycrystalline $(AgI)_{0.2}(Ag_2MoO_4)_{0.8}$ solid electrolyte [6], no RESET/erase mechanism works for the present case. Therefore, because of only write-read sweep, the switching behavior has been recognized as an irreversible memory type which is particularly important for non-erasable read-only memory technology. Figure 3c and 3d show the I-V characteristics of representative samples of 0.2 mm thickness.

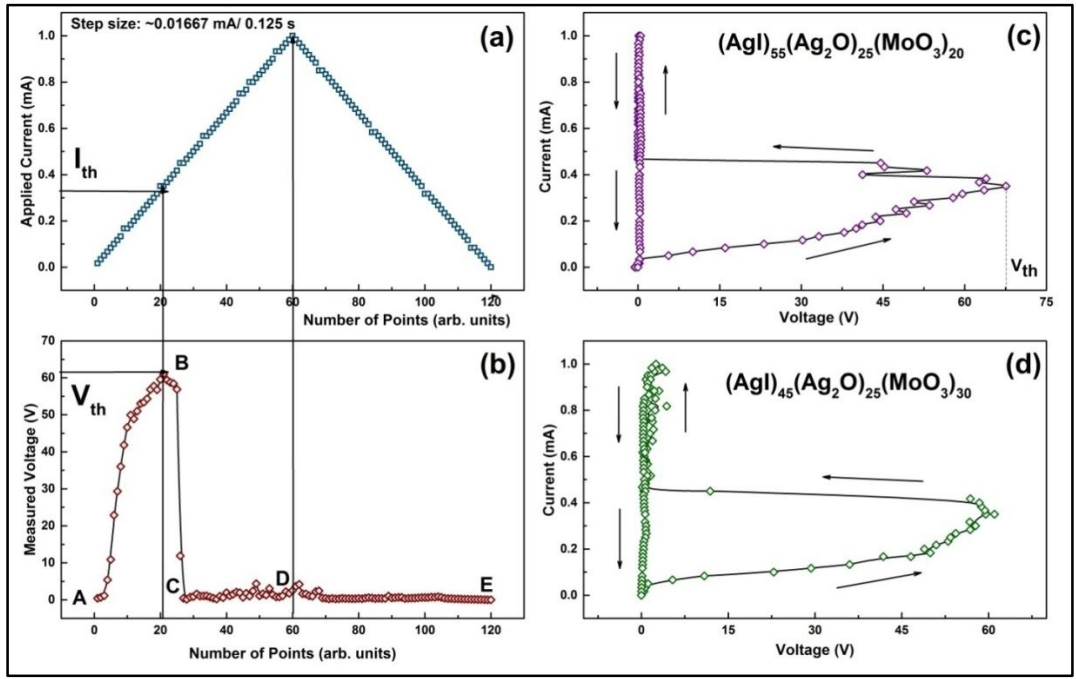

Figure 3: (a) Applied triangular current pulse (mA) with step size ~ 0.01667 mA/0.125 s (b) Measured voltage vs. Number of points (arbitrary units) (c) I-V characteristics of S3 - $(AgI)_{55}(Ag_2O)_{25}(MoO_3)_{20}$ (d) I-V characteristics of S7 - $(AgI)_{45}(Ag_2O)_{25}(MoO_3)_{30}$

## C. The Switching Mechanism

The switching phenomenon has been understood in the framework of cation based resistive switching mechanism [5, 25-27] that involves an electrochemical metallization process. Principally, the active anode oxidizes and produces cations ($M \rightarrow M^{z+} + ze^-$) that migrate across the solid electrolyte as a result of the applied electric field. This migration follows the Mott-Gurney model for electric field driven, thermally activated ion hopping [27]. The dependence of ionic current density (i) on electric field (E) is given by:

$$i = 2zec\alpha\vartheta\left[\exp\left(-\frac{W_a^0}{kT}\right)\right]\left[\sinh\left(\frac{\alpha zeE}{2kT}\right)\right] \tag{1}$$

Where c is the mobile cation ($M^{z+}$) concentration, $\alpha$ is jump distance of ions, $\nu$ is a frequency factor, k is Boltzmann's constant, T is temperature, $W_a^0$ is a symmetric energy barrier for thermally activated ion hopping; electric field, $E = \Delta\varphi_{SE}/d$; where $\Delta\varphi_{SE}$ is the potential drop across the electrolyte and d is the thickness.

As these cations reach to the inert cathode, cathodic deposition takes place ($M^{z+} + ze^- \rightarrow M$) and a successive coagulation of cations causes the metallic filament formation between two electrodes, inverting the state into low resistance state (LRS). The RESET occurs when the electrode polarity is altered and as a result, active electrode oxidation ceases to happen.

In this present study, point contact bronze electrodes are used; thus the anode oxidation and dissolution does not take place because of higher contact potential between electrode and electrolyte. EDS studies (Supplementary Figure 3 and 4) on the switched spot of a sample does not show any traces of constituents of bronze in the metallic filament. Thus, only ions within the solid-electrolyte take part in conduction and filament formation. The present electrolyte consists of $Ag^+$, $I^-$ and $[MoO_4]^{2-}$ ions [17, 28] and due to size constraints, only $Ag^+$ ions get involved in the conduction process.

As discussed in the sample preparation section, the solid structure constitutes of the glass former $MoO_3$ and network modifier $Ag_2O$ while the dopant ions ($Ag^+$ and $I^-$) do not contribute directly to the network formation but remain decoupled from the network, expand the network by creating free volume when introduced into the system[14, 29]. The $Ag^+$ ions transport due to applied electric field occurs through these free volumes. Once the cations get in contact with the inert cathode, metallization initiates as $Ag^+ + e^- \rightarrow Ag$. This metallic Ag keeps expanding through the free volume until the electrodes get shorted; in other words, the measured voltage profile continues from point A to B in Figure 3b, up to $V_{th}$ where the short happens. The I/V profile within A and B (Figure 3b), where the ions hop from site to site while driven by electric field, follows the Mott-Gurney model (Equation 1). The significance of fitting this specific segment of the curve with Mott-Gurney model lies in the understanding of the $V_{th}-x$ profile that will be discussed in the next section.

Once the filament is formed, it works like a unipolar fuse that remains unchanged with electrode polarity alteration. Furthermore, an effort to melt the filament thermally with RESET current impulse can deform the neighboring structure. So clearly, the electrode and contact types are so important that just by changing them from low work function active electrode [6] to a high contact potential inert electrode alters the memory type from reversible to irreversible.

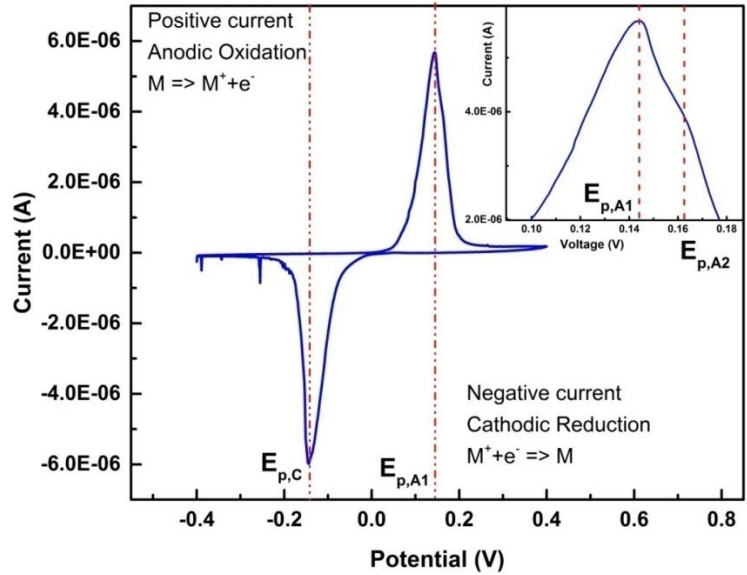

Figure4: Cyclic voltammogram of S3 - $(AgI)_{55}(Ag_2O)_{25}(MoO_3)_{20}$, with $E_{max} = 0.4$ V and $E_{min} = -0.4$ V, Scan rate = 0.1 V/s and sensitivity = 1E-5 A/V. (Inset) Close up for Anodic oxidation peak to identify a small protrusion as an latent process involving electron transfer kinetics

This metallization process has been distinguished from electrochemical polarization process. Figure 4 shows result obtained from CV measurement in a representative sample, S3 - $(AgI)_{55}(Ag_2O)_{25}(MoO_3)_{20}$ to understand this distinction. The current-voltage plot suggests that the electron transfer kinetics between electrode and electrolyte is electrochemically reversible. The reversibility criterions are [30, 31],

$$E_{P,A} - E_{P,C} = \frac{57.0}{n} \ mV \qquad (2)$$

$$\left| \frac{I_{P,A}}{I_{P,C}} \right| = 1 \qquad (3)$$

Where $E_P$ and $I_P$ are peak potential and current, A and C indices stand for anodic and cathodic processes respectively; n is the number of electrons participating in the electrochemical reactions. Present results show $E_{P,A1} = 0.145 \ V, E_{P,C} = -0.144 \ V$ , and $I_{P,A1} = 5.673E - 6 \ A, I_{P,C} = 5.987E - 6 \ A$ , i.e. the reversibility criterions satisfy within the error limit and for $n = 2$. Thus, the system attains Nernst's

Equilibrium swiftly after any change in applied potential. This swiftness restrains any possibility for polarization effect and hence filament formation due to this.

Moreover, a very small protrusion has been observed in the anodic process whose peak voltage has been denoted $E_{P,A2}$. This is significantly absent in cathodic process, implying a latent irreversible, corrosive reaction or oxidation near anode. This phenomenon will be discussed later in this work with the help of SEM images.

### D. $V_{th}$ and Composition

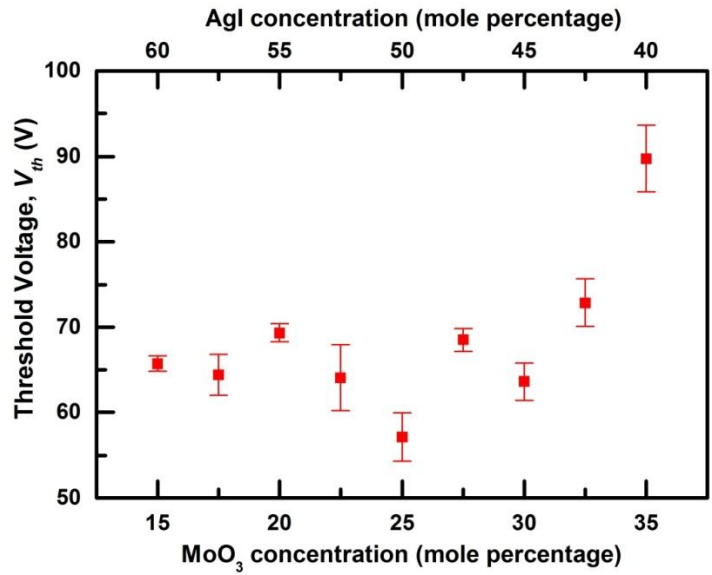

Figure 5: The variation of $V_{th}$ (V), with $MoO_3$ concentration (mole percentage)

Figure 5 shows the $V_{th}$ variation with $MoO_3$ (x) and AgI concentration. $V_{th}$ is a switching parameter used to establish a performance based classification of the materials. Faster the switching happens, smaller the value of $V_{th}$ becomes. Furthermore, the area under the I/V profile is directly proportional to the power loss during switching and $V_{th}$ is one of the measuring parameter of that area. The scattered nature of the plot confirms a loose correlation but there are two important aspects to describe the nature of this plot. The agility to reach $V_{th}$ would depend on the nature of Mott-Gurney model fitted segment of I/V profile in Figure 3 and hence, on the parameters that constitutes the ion transport mechanism. In Equation 1, the only composition related parameter is c, the mobile cation ($M^{z+}$) concentration. Interestingly, not all $Ag^+$ ions are mobile [28] and the mobile ion concentration depends on the environment surrounding the $Ag^+$ ions. Whereas, long range migration pathway requires a mixed iodine-oxygen coordination instead of an entire iodine environment because the pathways in this case gets restricted to very local region of few Å

[32-33]. Thus, for a particular sample, where there is balanced mixed iodine-oxygen coordination around $Ag^+$ ions, mobile ion concentration (c) increases. Now c plays a role of a multiplicative constant in the Mott-Gurney model of the form,

$$i = cA\{\sinh(BV)\} \tag{4}$$

Where $A = 2ze\alpha\vartheta[\exp(-W_a^0/kT)]$, $B = aez/2kdT$ and $V = \Delta\varphi_{SE}$. An increase in c would cause steepness rise in the I/V profile and hence a lower $V_{th}$. This happens to S5 - $(AgI)_{50}(Ag_2O)_{25}(MoO_3)_{25}$, the central composition, exhibits the lowest value of $V_{th}$.

The second aspect to understand the $V_{th}$–x profile comes from the context of its roughly increasing nature. $AgI$-$Ag_2O$-$MoO_3$ is a decoupled system i.e. in this solid electrolytic system; all the physical processes that are associated with the network structure and ion transport are decoupled [34]. Hence the increment of $V_{th}$, because of lessening of mobile ion concentration with higher $MoO_3$ concentration, is not as well defined as $T_g$– x profile in Figure 2. And the importance of this decoupled system lies in the fact that electrochemical applications in the decoupled system are possible below $T_g$ with significant cation mobility, that distinguishes it from coupled system [35]. A requirement for higher temperature to initiate electrochemical processes and hence, the electrical switching in a system would be costly and vulnerable to heat induced damages. Moreover, the thickness dependence of $V_{th}$ is also evident from Equation 4.

### E. SEM and Raman; Evidence of Anodic Oxidation

Microstructures of the surface of theun-switched sample (Figure 6a) and the vertical cross-section of the metallic filament of the switched samples (Figure 6b) are examined by scanning electron microscopy (SEM). The chemical nature of the filament and thebase surface, examined by EDS, shows around 73.66% increase in silver and 37.84% increase in iodine weight percentage (Wt. %) in the filament than the base material. No traces of electrode material have been found in the filament. Comparative normalized Raman spectrum of un-switched and the switched sample is shown in Figure 6c. Raman peaks for an un-switched sample are: $Ag^+$ restrahlen mode near 75 $cm^{-1}$[34], Ag–I mode at 95 $cm^{-1}$; $\nu_2$ (E), $\nu_4$ $(F_2)$, $\nu_3$ $(F_2)$ and $\nu_1$ $(A_1)$ modes for $MoO_4^{2-}$tetrahedra appears at 305 $cm^{-1}$, 332 $cm^{-1}$, 826 $cm^{-1}$ and 878 $cm^{-1}$ respectively; a new Mo–O bond stretching peak appears at 900 $cm^{-1}$, for a composition region where $Ag_2O/MoO_3 < 1$ [18]. After switching, coagulation of silver occurs due to filament formation which exhibits an intensity enhancement of Ag–I mode and a complete absence of Mo–O modes of molybdate. Figure 6e shows a schematic of the filament formation and the switching mechanism.

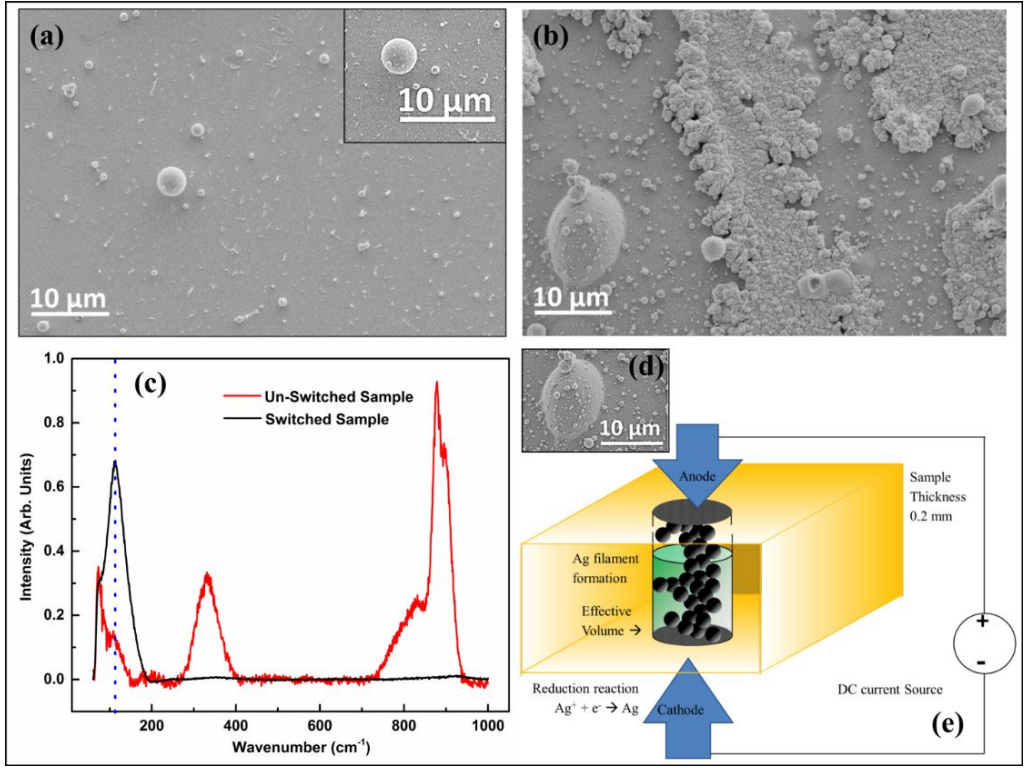

Figure 6: (a) SEM image of the surface of theun-switched sample. (b) SEM image of the same surface after switching; metallic filament formation between electrodes. (c) Normalized Raman spectrum of Un-switched and switched $(AgI)_{42.5}(Ag_2O)_{25}(MoO_3)_{32.5}$ sample. (d) Bubble formation, near Anode. (e) Schematic of the filament formation between two electrodes.

The globule and bubble formation is one of the most interesting results that we find in our SEM/EDS study. The globule in Figure 6(a. inset) is caused by substantial oxygen containment than the base material, 41.06% (Wt. %). After switching, bubbles form near anode, shown in Figure 6(d). A similar type of bubble formation due to surrounding partial pressure of oxygen and vacuum conditions has been observed in Au/SrTiO₃/Pt system where the switching mechanism is based on Oxygen transport and dislocations [36]. This evokes to reconsider the oxidation reaction in anode speculated earlier [3]. The oxidation reactions have been suggested as follows:

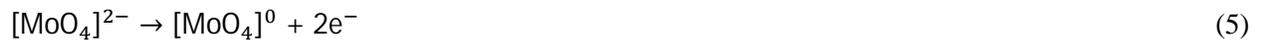

$$[MoO_4]^{2-} \rightarrow [MoO_4]^0 + 2e^- \tag{5}$$

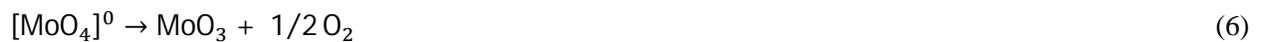

$$[MoO_4]^0 \rightarrow MoO_3 + 1/2\,O_2 \tag{6}$$

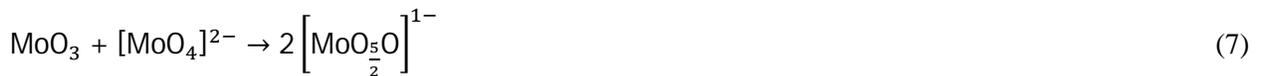

$$MoO_3 + [MoO_4]^{2-} \rightarrow 2\left[MoO_{\frac{5}{2}}O\right]^{1-} \tag{7}$$

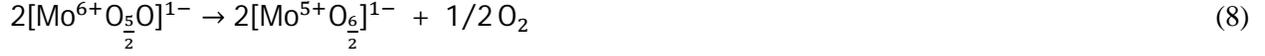

$$2[Mo^{6+}O_{\frac{5}{2}}O]^{1-} \rightarrow 2[Mo^{5+}O_{\frac{6}{2}}]^{1-} + 1/2 \, O_2 \hspace{2cm} (8)$$

According to the suggestions, the charge transport involves a reduction of $2Mo^{6+} \rightarrow 2Mo^{5+}$ and formation of extensive chains of $[Mo^{5+}O_{6/2}]^{1-}$ units; this clearly involves last two reactions (7 and 8). In the present study, the bubble formation occurs only in the vicinity of anode–electrolyte interface and not in the whole effective volume because there are no traces of excessive molybdenum or oxygen in the filament. Thus the extensive chain formation of $[Mo^{5+}O_{6/2}]^{1-}$ units, that causes the $2Mo^{6+} \rightarrow 2Mo^{5+}$ reduction is not happening. And hence, only first two reactions (5 and 6) are taking place where the molybdate ion $[MoO_4]^{2-}$ is being reduced to molybdenum tri-oxide and oxygen that results in bubble formation.

In case of MIM structured Ag (active)/Electrolyte/Pt (inert) system [5-6], oxidation reaction does not take place. However, as the electrode type is changed, oxidation takes place near anode along with the irreversible memory type switching. There are other important significances of the observed oxidation reaction; (I) it is an effective visual measure of corrosion in the material. The $[MoO_4]^{2-} \rightarrow MoO_3$ reduction directly affects the material structure (II) this confirms the earlier claim of oxidation reaction [3], though partially. The conduction process is primarily dominated by thermally activated cation hopping and no evidence of oxygen transport has been found. And hence the metallic path essentially constitutes silver, not molybdenum.

## 4. Conclusion

Swithing of bulk, FIC, AgI-Ag$_2$O-MoO$_3$ vitreous solids, within a thermally stable domain, with inert electrode exhibit fast, nearly ideal, irreversible memory type switching and very less power loss. The electrochemical process responsible for this switching involves reversible cathodic metallization and an irreversible anodic polarization that causes corrosion. The migration mechanism of ions and switching performance, in terms of agility to reach the threshold voltage, follows Mott-Gurney's model for electric field driven, thermally activated ion hopping. Traces of anionic transport are completely absent.

- **Acknowledgements**

We thank our past lab member Prof. M. Anbarasu (Electrical Engineering Department, IIT Indore) for the LabVIEW code for switching study. We thank Dr. Madhavi Kondia from Prof. G. Mohan Rao's Plasma Processing Lab (Instrumentation and Applied Physics, IISc Bangalore) for her very helpful assitance in Cyclic Voltammetry study. We also thank facility technologists Pradeep Kumar M. L. for Raman spectroscopy and Jose Martina for SEM/EDS (MNCF, CeNSE, IISC Bangalore). We are also thankful to the reviewrs for their very helpful and crucial comments.

**References**

[1] T. Minami, H. Nambu, and M. Tanaka, "Formation of Glasses with High Ionic Conductivity in the system AgI-Ag$_2$O-MoO$_3$," *J. Am. Ceram. Soc.*, **283** [June] 283–284 (1977).

[2] J. Kuwano, "Silver ion conducting glasses and some applications," *Solid State Ionics*, **40/41** 696–699 (1990).

[3] B. Vaidhyanathan, S. Asokan, and K.J. Rao, "Near Ideal electrical switching in fast ion conducting glasses: Evidence for an electronic process with chemical origin," *Bull. Mater. Sci.*, **18** [3] 301–307 (1995).

[4] B. Vaidhyanathan, K.J. Rao, S. Prakash, S. Murugavel, and S. Asokan, "Electrical switching in AgI based fast ion conducting glasses: Possibility for newer applications," *J. Appl. Phys.*, **78** 1358–1360 (1995).

[5] V. V. Kharton (ed.), *Solid state electrochemistry II: Electrodes, Interfaces and Ceramic Membranes*. Wiley-VCH Verlag GmbH & Co. KGaA, 2011.

[6] X.B. Yan, J. Yin, H.X. Guo, Y. Su, B. Xu, H.T. Li, D.W. Yan, Y.D. Xia, *et al.*, "Bipolar resistive switching performance of the nonvolatile memory cells based on (AgI)$_{0.2}$(Ag$_2$MoO$_4$)$_{0.8}$ solid electrolyte," *J. Appl. Phys.*, **106** 054501-1-054501-5 (2009).

[7] X.F. Liang, Y. Chen, L. Chen, J. Yin, and Z.G. Liu, "Electric switching and memory devices made

from RbAg$_4$I$_5$ films," *Appl. Phys. Lett.*, **90** [2] (2007).

[8]  T. Sakamoto, H. Sunamura, H. Kawaura, T. Hasegawa, T. Nakayama, and M. Aonob, "Nanometer-scale switches using copper sulfide," *Appl. Phys. Lett.*, **82** [18] 3032–3034 (2003).

[9]  K. Terabe, T. Nakayama, T. Hasegawa, and M. Aono, "Formation and disappearance of a nanoscale silver cluster realized by solid electrochemical reaction," *J. Appl. Phys.*, **91** [12] 10110–10114 (2002).

[10]  T. Takahashi, "Solid silver ion conductors," *J. Appl. Electrochem.*, **3** 79–90 (1973).

[11]  J.G.P. Binner, G. Dimitrakis, D.M. Price, M. Reading, and B. Vaidhyanathan, "Hysteresis in the β-α phase transition in Silver iodide," *J. Therm. Anal. Calorim.*, **84** 409–412 (2006).

[12]  G.R. Robb, A. Harrison, and A.G. Whittaker, "Temperature-resolved , in-situ powder X-ray diffraction of silver iodide under microwave irradiation," *Phys. Chem. Commun.*, **5** 135–137 (2002).

[13]  K.J. Rao, *Structural Chemistry of Glasses*, First. Elsevier Ltd., 2002.

[14]  J. Swenson and L. Börjesson, "Correlation between free volume and ionic conductivity in fast ion conducting glasses," *Phys. Rev. Lett.*, **77** [17] 3569–3572 (1996).

[15]  T. Minami, K. Imazawa, and M. Tanaka, "Formation region and characterization of superionic conducting glasses in the system AgI-Ag$_2$O-MoO$_3$," *J. Non. Cryst. Solids*, **42** 469–476 (1980).

[16]  T. Minami, "Fast ion conducting glasses," *J. Non. Cryst. Solids*, **73** 273–284 (1985).

[17]  T. Minami, T. Katsuda, and M. Tanaka, "Infrared Spectra and structure of superionic conducting glasses in the system AgI-Ag$_2$O-MoO$_3$," *J. Non. Cryst. Solids*, **29** 389–395 (1978).

[18]  N. Machida and H. Eckert, "FT-IR, FT-Raman and [95]Mo MAS–NMR studies on the structure of ionically conducting glasses in the system AgI–Ag$_2$O–MoO$_3$," *Solid State Ionics*, **107** 255–268 (1998).

[19]  V. V. Kharton (ed.), *Solid State Electrochemistry I: Fundamentals, Materials and their Applications*. Wiley-VCH Verlag GmbH & Co. KGaA, 2009.

[20]  M.H. Chiu and E.J. Prenner, "Differential scanning calorimetry: An invaluable tool for a detailed thermodynamic characterization of macromolecules and their interactions," *J. Pharm. Bioallied*

*Sci.*, **3** [1] 39–59 (2011).

21      E. D. Zanotto, J. C. Mauro, "The glassy state of matter: Its defination and ultimate fate," *J. Non. Cryst. Solids*, **490–495** 471 (2017).

22      P. Mustarelli, C. Tomasi, A. Magistris, and M. Cutroni, "Ion dynamics and devitrification in 0.75AgI-0.25Ag$_2$MoO$_4$ fast ion conducting glass: An XRD, DSC and $^{109}$Ag NMR study," *J. Non. Cryst. Solids*, **232–234** 532–538 (1998).

23      C. Tomasi, P. Mustarelli, and A. Magistris, "Devitrification and Metastability: Revisiting the Phase Diagram of the System AgI:Ag$_2$MoO$_4$," *J. Solid State Chem.*, **140** [1] 91–96 (1998).

24      G. Sreevidya Varma, D.V.S. Muthu, A.K. Sood, and S. Asokan, "Electrical switching, SET-RESET, and Raman scattering studies on Ge$_{15}$Te$_{80-x}$In$_5$Ag$_x$ glasses," *J. Appl. Phys.*, **115** 164505-1-164505–6 (2014).

25      T. Chang, K. Chang, T. Tsai, T. Chu, and S.M. Sze, "Resistance random access memory," *Mater. Today*, **19** [5] 254–264 (2016).

26      S. Tappertzhofen, I. Valov, T. Tsuruoka, T. Hesegawa, R. Waser, and M. Aono, "Generic Relevance of Counter Charges for Cation-Based Nanoscale Resistive Switching Memories," *ACS Nano*, **7** [7] 6396–6402 (2013).

27      B.R. Waser, R. Dittmann, G. Staikov, and K. Szot, "Redox-Based Resistive Switching Memories – Nanoionic Mechanisms , Prospects , and Challenges," *Adv. Mater.*, **21** 2632–2663 (2009).

28      T. Minami and M. Tanaka, "Structure and Ionic transport of Superionic Conducting Glasses in the System AgI-Ag$_2$O-MoO$_3$," *J. Non. Cryst. Solids*, **38&39** 289–294 (1980).

29      P. Mustarelli, C. Tomasi, and A. Magistris, "Fractal Nanochannels as the Basis of the Ionic Transport in AgI-Based Glasses," *J. Phys. Chem. B*, **109** 17417–17421 (2005).

30      C.M.A. Brett and A.M.O. Brett, *Electrochemistry: Principles, Methods and Applications*. Oxford University Press, 1994.

31      N. Elgrishi, K.J. Rountree, B.D. McCarthy, E.S. Rountree, T.T. Eisenhart, and J.L. Dempsey, "A Practical Beginner's Guide to Cyclic Voltammetry," *J. Chem. Educ.*, **95** [2] 197–206 (2018).

32      S. Adams and J. Swenson, "Migration pathways in Ag-based superionic glasses and crystals investigated by the bond valence method," *Phys. Rev. B*, **63** 054201-1-054201-11 (2000).

33    J. Swenson and S. Adams, "Application of the bond valence method to reverse Monte Carlo produced structural models of superionic glasses," *Phys. Rev. B*, **64** 024204-1-024204-10 (2001).

34    D.I. Novita, P. Boolchand, M. Malki, and M. Micoulaut, "Elastic flexibility , fast-ion conduction , boson and floppy modes in $AgPO_3$–$AgI$," *J. Phys. Condens. Matter*, **21** 205106-1-205106–17 (2009).

35    M.D. Ingram, C.T. Imrie, J. Ledru, and J.M. Hutchinson, "Unified Approach to Ion Transport and Structural Relaxation in Amorphous Polymers and Glasses," *J. Phys. Chem. B*, **112** 859–866 (2008).

36    K. Szot, W. Speier, G. Bihlmayer, and R. Waser, "Switching the electrical resistance of individual dislocations in single-crystalline $SrTiO_3$," *Nature*, **5** [April] 312–320 (2006).